
\magnification=\magstep1
\vsize=9truein
\hsize=6.5truein
\hoffset=0.25truein
\voffset=0.25truein
\overfullrule=0pt
\baselineskip=12pt
\centerline{\bf Comparison of Dynamical Approximation Schemes for}
\centerline{\bf Non--Linear Gravitational Clustering}
\bigskip
\centerline{\bf Adrian L. Melott}
\smallskip
\centerline{Department of Physics and Astronomy}
\centerline{University of Kansas}
\centerline{Lawrence, KS 66045}
\vskip .5in
\centerline{\bf ABSTRACT}
We have recently conducted a controlled comparison of a number of
approximations for gravitational clustering against the same $n$--body
simulations. These include ordinary linear perturbation theory (Eulerian),
the lognormal approximation,
the adhesion approximation, the frozen-flow approximation, the Zel'dovich
approximation (describable as first--order Lagrangian perturbation theory), and
its second--order generalization. In the last two cases we also created new
versions of the approximation by truncation, i.e., by smoothing the initial
conditions with various smoothing window shapes and varying their sizes.
\medskip
The primary tool for comparing simulations to approximation schemes was
crosscorrelation of the evolved mass density fields, testing the extent to
which
mass was moved to the right place. The Zel'dovich approximation, with initial
convolution with a Gaussian $e^{-k^2/k^2_G}$,  where $k_G$ is adjusted to be
just into the nonlinear regime of the evolved model (details in text) worked
extremely well. Its second--order generalization worked slightly better.
\medskip
All other schemes, including those proposed as generalizations of the
Zel'dovich approximation created by adding forces, were in fact generally
worse by this measure.
By explicitly checking, we verified that the success of our
best--choice was a result of the best treatment of the phases of nonlinear
Fourier components. Of all schemes tested, the adhesion approximation produced
the most accurate nonlinear power spectrum and density distribution, but its
phase errors suggest mass condensations were moved to
slightly the wrong location.
Due to its better reproduction of the mass density distribution function
and power spectrum, adhesion might be preferred for some uses.
\medskip
We recommend either $n$--body simulations or our modified versions
of the Zel'dovich approximation, depending upon the purpose. The theoretical
implication is that pancaking is implicit in all cosmological gravitational
clustering, at least from Gaussian initial conditions,
even when subcondensations are present.
This in turn provides a natural explanation for the presence of sheets and
filaments in the observed galaxy distribution. Use of the approximation scheme
can permit extremely rapid generation of large numbers of realizations of model
universes wtih good accuracy down to galaxy group mass scales.
\vfill
\vfill\eject
\noindent {\bf I. INTRODUCTION}
\medskip
The gravitational instability picture has emerged as the dominant paradigm for
understanding the growth of structure  in the universe from the
small--amplitude
fluctuations present at recombination. When the density fluctuations are very
small, linear perturbation theory (Eulerian) works well (for a summary see
Peebles 1980, 1993). In the deeply non--linear regime, $n$--body simulations
are usually used, perhaps with simulated hydrodynamics added.
Simulations generally suffer from the
typical fault of numerical results that they can be quite correct without
our understanding why.  This makes them difficult to generalize.
Even worse, without approximate analytic solutions as a check, errors
may be unrecognized.
Yet we need to be able to be surprised by the results
occaisionally, or we are not doing cutting-edge work.
\medskip
Analytic or quasi--analytic nonlinear approximations occupy an intermediate
position. They can capture some non--linear effects correctly in a way that
permits us to understand and generalize from them more easily. They can also be
used to provide boundary conditions for simulations, start them at a more
advanced state, or generate large numbers of approximate realizations, for
statistical purposes. It is for this reason that they are worth proposing; it
is
also for this reason that they are worth testing objectively in a way that
allows comparisons. In this {\it Letter} I report succinctly the main results
of
such a project.
\medskip
In Coles, Melott and Shandarin (1993), hereafter CMS, we began by studying the
usual linear (Eulerian) perturbation theory, the lognormal (Coles and
Jones, 1991) approximation (basically an exponentiation of linear) and the
Zel'dovich (1970) approximation. The lognormal approximation was particularly
poor and will not be considered further in this {\it Letter}.
\medskip
The Zel'dovich approximation (hereafter 1ZA) did very well, especially in a new
form in which the initial conditions were smoothed at a scale close to the
threshold of nonlinearity. Other schemes have been suggested recently
which are designed to be improvements on 1ZA.
\bigskip
\noindent {\bf II. APPROXIMATIONS STUDIED}
\medskip
Brief verbal descriptions will be presented for the many approximations tested.
For full details please see the citations.
\medskip
Linear theory results from perturbing the equations of
motion. The result is that (for $\Omega=1$, but easily generalized) the
fluctuations in merely grow in amplitude $\delta \propto a (t)$ where
$a$ is the scale factor. A velocity is of course also implied, but does not
produce the associated $\delta$ except to first order. See Peebles (1980,
1993).
\medskip
The Zel'dovich (1970) approximation (1ZA) consists of the assumption that the
velocity taken from linear theory continues. In comoving coordinates $\bar x$,
$\Omega=1$, this reduces to $d\bar x/a=\; constant$. The density is then
derived from the position of mass elements. The assumption that the velocity
will behave in this way seems most appropriate when the acceleration field is
constant over large regions, i.e. it is associated with
universes in which there is extensive damping of small-scale
density flucutations, such as those
with adiabatic baryon perturbations or hot dark matter. 1ZA was tested (up to
resolution limits) by CMS.
\medskip
Buchert (1992) provided the derivation that Zel'dovich never presented for
his approximation  as resulting from first--order perturbation theory around
the
Lagrangian equations of motion. Furthermore, he has extended them to second
order which includes tidal forces and even to third  order. We have studied 2ZA
and 3ZA as well for damped models. The 2ZA approach is being studied now
for the power--law spectra reported here
(Melott, Buchert, and Weiss, 1994).
\medskip
CMS introduced the ``truncated Zel'dovich approximation" (hereafter 1TZA) by
smoothing the initial conditions on the scale of nonlinearity before applying
the approximation. This consists of destroying information about deeply
nonlinear modes the approximation cannot handle. It was inspired by previous
observations that the pattern of arrangement of clumps in hierarchical
clustering simulations resembled those of the pancakes in simulations with the
same phases but all the initial power set to zero for modes that had gone
nonlinear (Melott and Shandarin 1990; Beacom {\it et al.} 1991; Little,
Weinberg and Park 1991; see also Melott and Shandarin 1993).
This suggests that long wave modes dominate the structural morphology and
behave according to the so-called pancake theory even when substructure
is present.  They found that
1TZA worked extremely well, outperforming everything else. More recently
Melott, Pellman and Shandarin (1994, hereafter MPS) searched for the optimum
scale and shape of truncation, resulting in considerable improvement over the
CMS formulation by using a Gaussian smoothing of the initial conditions. We
shall refer to this optimized form as 1TZA. It is important to stress that the
results of 1TZA do not resemble those of 1ZA for most spectra.
It is for this reason that claims that an approximation is better than
the Zel'dovich approximation (e.g. Bagla and Padmanabhan 1994) are not very
meaningful.  1ZA and 2ZA are
quite bad for power-law spectra $ n > -1 $ (see part III).
\medskip
The ``frozen flow" approximation (FFA) Matarresse {\it et al.} (1992) is one
in which the particles follow streamlines of the original linear velocity
field.
This takes multiple steps, because the particles' velocity depends on their
position. It is not strictly an analytic approximation but it is so fast we
will
include it.
FFA was tested by Lucchin {\it et al.} (1994).
\medskip
The adhesion approximation (AA) Gurbatov, Saichev, and Shandarin 1989) contains
ZA as a core, but adds an effective viscosity term which causes intersecting
flows to ``stick." This is an attempt to correct the first serious fault of ZA,
that particles continue past shell crossing in the approximation, but are
slowed
by gravity and fall back in the real world. Melott, Shandarin, and Weinberg
(1994) tested the adhesion approximation using the method of Weinberg and Gunn
(1990).
\medskip
Although plans exist to study it in the same way, we have not delayed this
letter to include results for the ``frozen potential" FP method (Brainerd {\it
et al.} 1994, Bagla and Padmanabhan 1994). Although it may perform well, it is
very far from an analytic approximation. It really consists of doing an
$n$--body simulation without re--solving for the potential
at each step.  This takes
advantage of the fact that the potential is constant to linear order, and
dominated by longwave modes which are little affected by nonlinearities. But
analytic solutions do not exist, and solving for the potential is easy with
modern numerical methods. It might provide insight, but it will never replace
an
analytic approximation {\it or} an $n$--body simulation. Perturbation theory
calculations (Munshi and Starobinsky 1993; Bernardeau {\it et al}. 1993)
suggest that FFA and FP may not be much improvement over ZA.
\medskip
One might propose generalizations of FFA, AA, or FP in which the initial
potential is smoothed. But in this case they would revert to something very
close to 1TZA. The whole purpose for creating them was to handle the nonlinear
modes, which 1TZA simply removes from the intial conditions.
\medskip
When trying to restore initial conditions from the evolved state, conclusions
presented herein do not apply. See Melott (1993).
\bigskip
\noindent {\bf III. PROCEDURES}
\medskip
All of our approximations are compared to a group of $n$--body simulations more
fully described in Melott and Shandarin (1993). These were $128^3$ particle
runs
with Gaussian initial conditions characterized by power--law spectra of
density fluctuations (see Peebles 1980) $P(k)\propto k^n$ for $n=-2,-1,0,+1$
which brackets most cases of cosmological interest. Conclusions about likely
behavior under specific physical scenarios can be reconstructed from the
power--law slopes just going nonlinear at the moment under consideration. The
$n$--body simulation and the approximations were compared primarily by
cross--correlation
$$S={<\delta_1\delta_2>\over \sigma_1\sigma_2}\; \; ,\eqno (1)$$ where
$\delta_i=<\delta_i^2>^{1/2}$,  $\delta_i$ is the pixellated density of the
simulation or approximation and $\sigma_i=<\delta_i^2>^{1/2}$. If they are
identical, $S=+1$. We allowed for the
fact that condensations might be just slightly in the wrong place by
calculating $S$ for both fields with a wide variety of identical Gaussian
smoothing lengths.
Some statistical analysis was also done, including power spectra and density
distribution functions. These will not be shown here, but can be found in the
various more detailed studies.
\medskip
The approach used here has a number of advantages over those used for testing
in most of
the approximation proposals. Most obviously, they are all tested against
the same initial conditions with the same methods, so they can be compared with
one another. We have also checked for detailed dynamical agreement rather than
just a similar visual appearance or power spectrum. One of the things we
learned
was that power spectra can be similar for two approximations while {\it phases}
can be in much
better agreement in one scheme than in another. Crosscorrelation is sensitive
to phase information.

\bigskip
\noindent {\bf IV. RESULTS}
\medskip
In Figure 1 I show the crosscorrelation $S$ as a function of $\sigma_1$ (the
$rms$ of the simulation),
for two indices $n=-1$ and
$+1$. The case $+1$ is the most demanding and shows the differences. The case
$-1$ is of interest because it is probably close to the slope just going
nonlinear today.
\medskip
It is clear that TZA is the best choice (for all mass scales) by this
criterion. 1TZA, or just TZA as
described by MPS, consists of Gaussian smoothing near the scale of
nonlinearity. We define $k_{n\ell}$ by
$$a^2(t)\int^{k_{n\ell}}_0P(k)d^3k\equiv 1\eqno (2)$$ where $P$ is the power in
the initial conditions. Therefore $k_{n\ell}$ is the wavenumber where
$\sigma=1$
by extrapolation using linear theory. MPS found that the optimum smoothing was
convolution of initial density by a Gaussian $e^{-k^2/2k_G^2}$ with $k_G=1.5\;
k_{n\ell} (n=-2,-1),\; 1.25 k_{n\ell}(n=0)$ or $k_{n\ell}(n=+1)$. As the
maximum is fairly broad, one could use $k_G=1.25k_{n\ell}$ for all cases
without serious error. In the case of non--power law spectra we recommend
examining the local slope at $k_{n\ell}$.
\medskip
If one wishes to perform a realization of a specific scenario he should
determine $k_{n\ell}$ from linear theory given the amplitude normalization
desired, then perform any biasing desired. For $b=1$, the scaling results
described in Melott and Shandarin (1993), section 6, imply that initial
conditions should be convolved with a Gaussian $e^{-R^2/R_G^2}$, where $R_G\sim
3.5 h^{-1}$ Mpc, before application of the Zel'dovich approximation, for
optimum results. This is a galaxy group mass scale, suggesting we can follow
things to that scale with this approximation.
\medskip
The Zel'dovich approximation is well known to produce ``pancakes", and its
agreement with a wide variety of simulated spectra implies that the pancaking
process is
a generic part of gravitational clustering (see also Dubinski {\it et al}.
1993). Figures 2 and 3 allow us to see the tracery of hidden pancakes in the
$n$--body simulations and also provide an idea of the level of detail likely to
be missed by our recommended approximation.
\medskip
This approximation scheme is within reach of anyone who has code to implement
the
Zel'dovich approximation, and a Fast Fourier Transform. It is extremely simple
to implement, and takes about as much CPU time as one step in an $n$--body
simulation. MPS also checked for agreement of particle positions and
velocities,
and found generally small errors.
Borgani et al (1994) have already used 1TZA to generate large ensembles
to test cluster--cluster correlations.
\medskip
Preliminary results indicate that 2TZA (Second order Lagrangian perturbation
theory with Gaussian smoothing of initial conditions) is a small but
measurable improvement over 1TZA (Melott, Buchert, and Weiss 1994).
It is important to also note that as reported in detail
by Melott, Shandarin, and Weinberg (1994), the adhesion approximation
more accurately reproduces the power spectrum and mass density distribution
function of the simulations than does 1TZA.
For spectral index $ n < -1$, the crosscorrelation is not bad, and for
this reason the adhesion approximation might be preferred for certain
purposes.
For most purposes, use of 1TZA is a simply implemented major improvement
over approximations now in use.
\medskip
The validity of this approximation is strong evidence that the existence of
sheets and filaments in the galaxy distribution can be a natural result of the
action of gravity on small--amplitude Gaussian initial conditions.
\vfill\eject
\noindent {\bf V. ACKNOWLEDGEMENTS}
\medskip
I thank in advance
the Aspen Center for Physics, which has graciously agreed to sponsor a workshop
on this and closely related topics in June 1994.
This research was supported in the USA by NASA (NAGW--2923) and NSF
(AST--9021414). For my collaborators it was also supported by (in the US)
the W.M. Keck Foundation and NSF grant PHYS 92--45317. It was supported
in Germany by DFG. In Italy we thank MURST and the CINECA Computing Center.
Primary production of simulations was done on a Cray--2 and Convex C3 at the
National center for Supercomputing Applications, Urbana, IL.
\vfill\eject
\centerline{\bf REFERENCES}
\medskip
\def\ref{\par\noindent\hangindent\parindent\hangafter1}
\ref
Bagla, J.S. and Padmanabhan, T. 1994, MNRAS, 266, 227.
\medskip
\ref
Beacom, J.F., Dominik, K.G., Melott, A.K., Perkins, S.F. and Shandarin,
S.F. 1991, ApJ 372, 351.
\medskip
\ref
Bernardeau, F., Singh, T.P., Banerjee, B., and Chitre, S.M. 1993 preprint.
\medskip
\ref
Borgani, S., Coles, P., and Moscardini, L. 1994, MNRAS, in press.
\medskip
\ref
Brainerd, T.G., Scherrer, R.J. and Villumsen, J.V. 1993, ApJ, 418, 570.
\medskip
\ref
Buchert, T. 1992, MNRAS, 254, 729.
\medskip
\ref
Coles, P. and Jones, B.J.T. 1991, MNRAS, 248, 1.
\medskip
\ref
Coles, P., Melott, A.L., and Shandarin, S.F. 1993, MNRAS, 260, 765.
\medskip
\ref
Dubinski, J., daCosta, L.N., Goldwirth, D.S., Lecar, M. and Piran, T. 1993, ApJ
410, 458.
\medskip
\ref
Gurbatov, S.N., Saichev, A.T. and Shandarin, S.F. 1989, MNRAS 236, 385.
\medskip
\ref
Little, B., Weinberg, D.H., and Park, C.B. 1991, MNRAS 253, 295.
\medskip
\ref
Lucchin, F., Matarrese, S., Melott, A.L. and Moscardini, L. 1994, MNRAS, in
press.
\medskip
\ref
Matarrese, S., Lucchin, F., Moscardini, L., and Saez, D. 1992, MNRAS 259, 437.
\medskip
\ref
Melott, A.L. 1993, ApJ Lett 414, L73.
\medskip
\ref
Melott, A.L., Buchert, T. and Weiss, A. 1994, Astron Ap, in preparation.
\medskip
\ref
Melott, A.L., Pellman, T.F. and Shandarin, S.F. 1993, MNRAS, submitted.
\medskip
\ref
Melott, A.L. and Shandarin, S.F. 1990, Nature 346, 633.
\medskip
\ref
Melott, A.L. and Shandarin, S.F. 1993, ApJ 410, 469.
\medskip
\ref
Melott, A.L., Shandarin, S.F., and Weinberg, D.H. 1994, ApJ, in press.
\medskip
\ref
Munshi, D. and Starobinsky, A.A. 1993 preprint.
\medskip
\ref
Peebles, P.J.E. 1980 {\it The Large--Scale Structure of the Universe}
(Princeton:Princeton University Press).
\medskip
\ref
Peebles, P.J.E. 1993 {\it Principles of Physical Cosmology}
(Princeton:Princeton University Press).
\medskip
\ref
Weinberg, D.H. and Gunn, J.E. 1990, MNRAS 247, 260.
\medskip
\ref
Zel'dovich, Ya.B. 1970, Astron Ap 5, 84.
\vfill\eject
\centerline{\bf FIGURE CAPTIONS}
\bigskip
\ref
Figure 1 A plot of the crosscorrelation $S$ of each of the various approximate
solutions with the $n$body simulation, both being
smoothed by the same
(variable) size Gaussian window, against $\sigma$, the $rms$ density
fluctuation in the smoothed $n$--body simulation. Results are shown for
spectral indices $n=+1$ and $n=-1$ at the moment when $k_{n\ell}=8k_f$,
where $k_f$ is the
fundamental mode of the box. In order of increasing accuracy, linear theory is
the short dashed line, the frozen flow approximation is the dottled line, the
adhesion approximation is the long dashed line, and the truncated Zel'dovich
approximation (1TZA) is the solid line.
\medskip
\ref
Figure 2 (a) A greyscale plot of a thin slice of the $n=+1$ $n$body simulation
at
the moment when $k_{n\ell}=8k_f$. (b) A corresponding slice of the 1TZA
approximation to the same.
\medskip
\ref
Figure 3 (a) As in Figure 2(a) but for $n=-1$. (b) As in Figure 2(b), but for
$n=-1$.
\bye